\documentstyle[prl,twocolumn,aps,epsf]{revtex}

\tighten

\begin{document}
\draft

\twocolumn[\hsize\textwidth\columnwidth\hsize\csname @twocolumnfalse\endcsname
\title{Trapping of single atoms in cavity QED}

\author{J. Ye, D. W. Vernooy, and H.J. Kimble}
\address{Norman Bridge Laboratory of Physics, California\\
Institute of Technology 12-33, Pasadena, CA 91125}
\date{\today}
\maketitle
\begin{abstract}
By integrating the techniques of laser cooling and trapping with those of
cavity quantum electrodynamics (QED), single Cesium atoms have been trapped
within the mode of a small, high finesse optical cavity in a regime of
strong coupling. The observed lifetime for individual atoms trapped within
the cavity mode is $\tau \approx 28$ms, and is limited by fluctuations of
light forces arising from the far-detuned intracavity field. This initial realization of
trapped atoms in cavity QED should enable diverse protocols in quantum
information science.
\end{abstract}
\vspace{3ex}
]

Cavity quantum electrodynamics (QED) offers powerful possibilities for the
deterministic control of atom-photon interactions quantum by quantum.\cite
{berman94,nobel} Indeed, modern experiments in cavity QED have achieved the
exceptional circumstance of strong coupling, for which single quanta can
profoundly impact the dynamics of the atom-cavity system. Cavity QED has led
to many new phenomena, including the realization of a quantum phase gate,
\cite{turchette95} the creation of Fock states of the radiation field,\cite
{walther99} and the demonstration of quantum nondemolition detection for
single photons\cite{haroche99}.

These and other diverse accomplishments set the stage for advances into yet
broader frontiers in quantum information science for which cavity QED offers
unique advantages. For example, it should be possible to realize complex
quantum circuits and quantum networks by way of multiple atom-cavity systems
linked by optical interconnects,\cite{pellizzari95,cirac98} as well as to
pursue more general investigations of quantum dynamics for continuously
observed open quantum systems\cite{mabuchi-wiseman}. The primary technical
challenge on the road toward these scientific goals is the need to trap and
localize atoms within a cavity in a setting suitable for strong coupling. In
fact, all proposed schemes for quantum computation and communication via
cavity QED rely implicitly on the development of techniques for atom
confinement that do not interfere with cavity QED interactions.

In this Letter we report a significant milestone in this quest, namely the
first trapping of a single atom in cavity QED. Our experiment integrates the
techniques of laser cooling and trapping with those of cavity QED to deliver
cold atoms (kinetic energy $E_{k}\simeq 30\mu $K) into the mode of a high
finesse optical cavity. In a domain of strong coupling, the trajectory of an
individual atom within the cavity mode can be monitored in {\it real time}
by a near resonant field with mean intracavity photon number $\bar{n}<1$\cite
{mabuchi96,hood98,mabuchi99,ye99,rempe99}. Here we exploit this capability
to trigger {\it ON} an auxiliary field that functions as a far-off resonance
dipole-force trap (FORT)\cite{henfort,chufort}, providing a confining
potential to trap the atom within the cavity mode. Likewise, when the FORT\
is turned {\it OFF} after a variable delay, strong coupling enables
detection of the atom. Repetition of such measurements yield a trap lifetime
$\tau =(28\pm 6)$ms, which is currently limited by fluctuations in the
intensity of the intracavity trapping field (FORT).

Stated in units of the coupling parameter $g_{0}$(where $2g_{0}$ is the
single-photon Rabi frequency), our work achieves $g_{0}\tau \simeq 10^{6}\pi
$, whereas prior experiments with cold atoms have attained $g_{0}T\simeq
10^{4}\pi $\cite{mabuchi96,hood98,mabuchi99,ye99,rempe99} and experiments
with conventional atomic beams have $g_{0}T\simeq \pi $,\cite
{berman94,nobel,turchette95,walther99,haroche99} with $T$ as the atomic
transit time through the cavity mode. Finally, as a step toward {\it in situ}
monitoring of an atom within the FORT we describe observations of the
transmission of a cavity QED probe field in the presence of the trapping
potential for single atom transits.

Our experimental apparatus consists of a high finesse cavity, two-stage
magneto-optical traps (MOT), and cavity probe, lock, and FORT beams, as
shown in Figure 1. Roughly $10^{8}$ Cesium atoms are accumulated in an
``upstairs'' MOT-$1$, cooled with polarization gradients to $3\mu $K, and
then transferred with $10\%$ efficiency to a ``downstairs'' MOT-$2$ located
in a UHV chamber with background pressure $10^{-10}$Torr. The captured atoms
are next cooled to $2\mu $K and dropped from a position $5$mm above a high
finesse optical cavity. Some few atoms fall between the cavity mirrors, and
thence through the cavity mode itself.

A final stage in the protocol for delivering cold atoms into the mode volume
is provided by a set of cooling beams located in the $y-z$ plane
perpendicular to the cavity axis, as shown in Fig. 1 (b). These beams form
two independent standing waves along the $\pm 45^{\circ }$ directions in the
$y-z$ plane, each with helical polarization, and are switched on for $1.5$ms
to remove the residual fall velocity of atoms arriving at the cavity mode
from MOT-$2$, leading to final velocities $v\sim 5$cm/s for atoms in the
immediate vicinity of the cavity mode.

The Fabry-Perot cavity into which the atoms fall is formed from two
super-polished spherical mirrors. The cavity length $l=44.6\mu $m, waist $%
w_{0}=20\mu $m, and finesse ${\cal F}=4.2\times 10^{5}$, and hence a cavity
field decay rate $\kappa /2\pi =4$MHz\cite{birefringence}. The atomic
transition employed for cavity QED is the $(g\equiv
6S_{1/2},F=4,m_{F}=4\rightarrow e\equiv 6P_{3/2},F=5,m_{F}=5)$ component of
the $D_{2}$ line of atomic Cesium\ at $\lambda _{atom}\equiv c/\nu
_{atom}=852.4$nm. For our cavity geometry and from the atomic transition
properties, we have $(g_{0},\gamma _{\bot })/2\pi =(32,2.6)$MHz, with $g_{0}$
as the peak atom-field coupling coefficient and $\gamma _{\bot }$ as the
dipole decay rate for the $e\rightarrow g$ transition. These rates lead to
critical photon and atom numbers $(m_{0}\equiv \gamma _{\bot
}^{2}/2g_{0}^{2},N_{0}\equiv 2\kappa \gamma _{\bot }/g_{0}^{2})=(0.003,0.02)$%
.

The cavity length is stabilized with an auxiliary diode laser $\lambda
_{lock}\equiv c/\nu _{lock}\approx 836$nm, which is stabilized relative to $%
\nu _{atom}$ by way of an auxiliary ``transfer cavity''\cite{mabuchi99}. It
is detuned $2$ longitudinal-mode orders {\it above} the cavity QED\ mode at $%
\nu _{cavity}\approx \nu _{atom}$, and creates a small AC Stark shift of $50$%
kHz in $\nu _{atom}$. Residual fluctuations in the length of the locked
cavity lead to variations in $\Delta _{ac}\equiv \nu _{atom}-\nu _{cavity}$
of $\delta \Delta _{ac}\approx \pm 10$kHz contained within a locking
bandwidth of about $10$kHz.

The ``trajectory'' of an individual atom is monitored in real time as it
enters and moves within the cavity mode by recording modifications of the
(pW-scale) cavity transmission for a circularly polarized probe field ${\cal %
E}_{probe}$ of frequency $\nu _{probe}=\nu _{atom}+\Delta _{probe}$. For our
Fabry-Perot cavity, the spatially dependent coupling coefficient $g(\vec{r}%
)=g_{0}\sin (2\pi x/\lambda _{cavity})\exp (-(y^{2}+z^{2})/w_{0}^{2})\equiv
g_{0}\psi (\vec{r},\lambda _{cavity})$, with the mirrors located at $x=(0,l)$%
. Heterodyne detection of the transmitted probe (with overall efficiency
47\%) allows inference of the atomic position in a fashion that can be close
to the standard quantum limit\cite{mabuchi99}.

For the purpose of atomic trapping, the transmitted probe beam can be
employed to trigger {\it ON} a far-off-resonance trap (FORT) \cite
{henfort,chufort} given the detection of an atom entering the mode volume.
Here, the FORT beam is derived from an external diode laser locked to a
cavity mode at $\lambda _{FORT}=c/\nu _{FORT}=869$nm, two longitudinal-mode
orders {\it below} the cavity QED mode at $\nu _{cavity}$. In this case, the
standing-wave patterns of the two modes at $(\nu _{FORT},\nu _{cavity})$ are
such that there are approximately coincident antinodes near the center and
ends of the cavity; hence, the trapping potential of the FORT has maximum
depths at the positions of maxima $\left( g_{0}\right) $ for cavity QED
coupling in these regions.

An example of the trapping of a single atom is given in Figure 2. In (a) the
arrival of an atom is sensed by a reduction in transmission for the probe
beam (of photon number $\bar{n}\approx 0.1$\cite{nbar}). The falling edge of
the probe transmission triggers {\it ON} the FORT field, which then remains
on until being switched {\it OFF} after a fixed interval. The presence of
the atom at this {\it OFF} time is likewise detected by modification of the
probe transmission, demonstrating a trapping time of $13.5$ms for the
particular event shown in the figure. Note that because the probabilities
for atom trapping given a trigger $p_{tp|tg}$ and for detection given a
trapped atom $p_{d|tp}$ are rather small ( $p_{tp|tg}p_{d|tp}\sim 0.03$), we
operate at rather high densities of cold atoms, such that the average atom
number present in the cavity mode at the time of the trigger is $\bar{N}%
_{atom}\sim 0.5$ (but which then falls off rapidly). As a consequence, the
atom that causes the trigger is not always the atom that is actually trapped
when the FORT is gated {\it ON}, with such `phantom' events estimated to
occur in roughly $1$ of $4$ cases.

The timing diagram for switching of the various fields is given in Figure
2(b). Note that although the probe field is left on for all times in Fig.
2(a), there is no apparent change in cavity transmission during the interval
in which an atom is purportedly trapped within the cavity mode. The absence
of atomic signatures during the trapping time, but not before or after, is
due to AC-Stark shifts associated with the FORT and/or the mismatched
antinodes between $(\nu _{FORT},\nu _{cavity})$. For the data of Figure 2, a
power of $30\mu $W incident upon the cavity at $\lambda _{FORT}$ leads to a
circulating intracavity power of $1$W, and to AC-Stark shifts $\Delta
_{FORT}^{e,g}=\pm 45$MHz for the excited $e$ and ground $g$ states at the
cavity antinodes, so that at these locations, the net atomic transition
frequency $\nu _{atom}$\ is shifted to the blue by $\Delta
_{FORT}^{e}-\Delta _{FORT}^{g}\equiv \Delta _{FORT}=+90$MHz. Moreover, the
spatial dependence of the cavity mode means that $\Delta _{FORT}^{e,g}(\vec{r%
})=\Delta _{FORT}^{e,g}\psi (\vec{r},\lambda _{FORT})$, so that the FORT
effectively provides a spatially dependent detuning that shifts the cavity
QED interactions out of resonance, with $\Delta _{ac}\rightarrow \Delta
_{ac}+\Delta _{FORT}(\vec{r})$. To calculate the probe transmission in this
case requires an analysis of the eigenvalue structure incorporating both the
coupling $g_{0}(\vec{r})$ as well as $\Delta _{FORT}(\vec{r})$\cite
{vernooy97}.

To avoid questions related to the complexity of this eigenvalue structure as
well as to possible heating or cooling by the probe field, we synchronously
gate {\it OFF} the probe field ${\cal E}_{probe}$ for measurements of trap
lifetime, with the result displayed in Figure 3. These data are acquired for
repeated trials as in Fig. 2 (namely, with the presence of an atom used to
trigger {\it ON} the FORT now of depth $\Delta _{FORT}^{g}=-50$MHz), but now
with the probe field gated {\it OFF} after receipt of a valid trigger (as
shown by the solid trace for ${\cal E}_{probe}$ in Fig. 2(b)). At the end of
the trapping interval, ${\cal E}_{probe}$ is gated back {\it ON}, and the
success (or failure) of atomic detection recorded. The lifetime for single
atoms trapped within the FORT is thereby determined to be $\tau
_{FORT}=(28\pm 6)$ms. This trap lifetime is confirmed in an independent
experiment where the FORT is turned on and off at predetermined times
without transit-triggering, yielding $\tau _{FORT}^{^{\prime }}=(27\pm 6)$%
ms. As mentioned in the discussion of Fig. 2, our ability to load the trap
with reasonable efficiency via asynchronous turn-on is due to operation with
large $\bar{N}_{atom}$.

Note that at each of the time delays in Fig. 3, a subtraction of
``background'' events (atomic transits delayed by the intracavity cooling
beams) has been made from the set of total detected events. We determine
this background by way of measurements following the same protocol as in
Fig. 2(b), but without the FORT beam. For times below $10$ms in Fig. 3, this
background dominates the signal by roughly $50$-fold, precluding accurate
measurements of trapped events. However, because it has a rapid decay time $%
\approx 3$ms, for times greater than about $20$ms it makes a negligible
contribution.

As for the factors that limit the trap lifetime, the spontaneous photon
scattering rate is 37 $s^{-1}$in our FORT. The trap lifetime set by
background gas collisions at a pressure of $10^{-10}$Torr is estimated to be
$\sim 100$s, which is likewise much longer than that actually observed.
However, Savard et al. \cite{thomas97} have shown that laser intensity noise
causes heating in a FORT with heating rate $\tau _{e}^{-1}=\pi ^{2}\nu
_{tr}^{2}S_{e}(2\nu _{tr})$ (Eq. (12) of Ref.\cite{thomas97}). Here $\nu
_{tr}$ is the trap oscillation frequency (in cycles/s) and $S_{e}(2\nu
_{tr}) $ is power spectral density of fractional intensity noise evaluated
at frequency $2\nu _{tr}$. For the FORT of Figure 3, we estimate $(\nu
_{tr}^{radial},\nu _{tr}^{axial})\approx (5,450)$kHz for the radial $(y,z)$
and axial $x$ directions, respectively. Direct measurements of the spectral
density of photocurrent fluctuations for the FORT beam emerging from the
cavity (calibrated by coherent AM at the requisite frequency $2\nu _{tr}$)
lead to $(S_{e}(2\nu _{tr}^{radial}),S_{e}(2\nu _{tr}^{axial}))\approx
(5\times 10^{-9},2.3\times 10^{-11})/$Hz, so that $(\tau _{e}^{radial},\tau
_{e}^{axial})\approx (830,23)$ms. The heating rate for $1/\tau _{e}^{axial}$
is in reasonable agreement with the observed $(1/e)$ trap decay rate $1/\tau
_{FORT}\simeq 1/28$ms, leading to the conclusion that fluctuations in
intracavity intensity drive heating along the cavity axis and are the
limiting factor in our current work. Such fluctuations are exacerbated by
the conversion of FM to AM noise of the FORT laser due to the high cavity
finesse at the wavelength of the FORT (here, ${\cal F}_{FORT}=3.5\times
10^{5}$). An avenue to reduce $S_{e}(2\nu _{tr}^{axial})$ is by way of
active servo control of the intensity of the intracavity FORT field (in
addition to a wider bandwidth frequency servo that keeps the FORT field
locked to the cavity resonance), which is a strategy that we are pursuing.

Finally, we return to the more general question of cavity QED in the
presence of the FORT. As a starting point in a more complete investigation,
Figure 4 displays a series of four atomic transits, each of increasing
duration. With the FORT {\it OFF}, the ``down-going'' transit in (a) arises
from an atom that was dropped from MOT-$2$ without the application of the
cooling pulse shown in Fig. 2(b) and provides a reference for the time of
free fall through $\psi (\vec{r},\lambda _{cavity})$ (here, $T\approx 100\mu
$s for $v\approx 30$cm/s). By contrast, with the cooling pulse applied (but
with the FORT\ still {\it OFF}), the transit in (b) is lengthened to $%
T\approx 420\mu $s. In (c), $\Delta _{probe}$ is altered to sense
``up-going'' transits, with now $T\approx 1$ms. Because the kinetic energy
of an atom with $v\sim 5$cm/s is much smaller than the coherent coupling
energy $\hbar g_{0}$, it is possible to achieve long localization times via
the single-photon trapping and cooling mechanisms discussed in Refs. \cite
{doherty,ritsch}, which can be understood by way of a simple
`Sisyphus'-picture based upon the spatially dependent level structure in
cavity QED. For the last trace in (d), the FORT\ is always {\it ON} (i.e.,
not gated as in Fig. 2), but with a shallower potential ($\Delta
_{FORT}^{g}=-15$MHz) than that in Fig. 2(a). We select the detunings $\Delta
_{probe}=-10$MHz and $\Delta _{ac}=-10$MHz to enhance observation of a
trapped atom via the composite eigenvalue structure associated with $g(\vec{r%
})$ and $\Delta _{FORT}(\vec{r})$\cite{eigenvalues}. We also expect that
cavity-assisted Sisyphus cooling \cite{ritsch} should be effective in this
setting. As in (d), this results in remarkable `transits' observed in
real-time with $T\approx 7$ms, corresponding to transit velocity $\bar{v}%
\equiv \frac{2w_{0}}{T}\approx 6$mm/s and associated kinetic energy $\frac{1%
}{2}m\bar{v}^{2}\sim h\nu _{tr}^{radial}\ll h\nu _{tr}^{axial}$.

In conclusion, although these are encouraging first results for trapping of
single atoms in cavity QED, an outstanding problem with dipole-force traps
is that the excited state experiences a positive AC Stark shift, leading to
an excited state atom being {\it repelled} from the trap (e.g., during
quantum logic operations). As well, the effective detuning $\Delta _{ac}(%
\vec{r})\equiv \Delta _{ac}+\Delta _{FORT}(\vec{r})$ is a strong function of
the atom's position within the trap. Fortunately, it turns out that a
judicious choice of $\lambda _{FORT}$ can eliminate both of these problems
by making $\Delta _{FORT}^{e}(\vec{r})=\Delta _{FORT}^{g}(\vec{r})<0$, and
hence $\Delta _{FORT}(\vec{r})=0$\cite{hood-wood99}. Alternatively, even for
the current setup, it should be possible to tune $\Delta _{FORT}^{e}$
together with $\Delta _{ac}$ to produce regions within the cavity mode for
which the spatially dependent level shift of a composite dressed state in
the first excited manifold matches $\Delta _{FORT}^{g}(\vec{r})$ for the
(trapping) ground state\cite{eigenvalues}, as was attempted in Fig. 4 (d).
These schemes in concert with extensions of the capabilities presented in
this Letter should allow us to achieve atomic confinement in the Lamb-Dicke
regime (i.e., $\eta _{x}\equiv 2\pi \Delta x/\lambda \ll 1$) in a setting
for which the trapping potential for the atomic center-of-mass motion is
independent of internal atomic state, as has been so powerfully exploited
with trapped ions\cite{wineland}. Generally speaking, this essential task
must be completed for long-term progress in quantum information science via
photon-atom interactions.

We enthusiastically acknowledge the contributions of C. J. Hood, T. W. Lynn,
and H. Mabuchi. This work was funded by the NSF, by DARPA via the QUIC
(Quantum Information and Computing) program administered by ARO, and by the
ONR. JY is supported by a Millikan Prize Postdoctoral Fellowship.

\begin{figure}[htbp]
\label{}
  \begin{center}
   \leavevmode
     \epsfxsize=7.2cm  \epsfbox{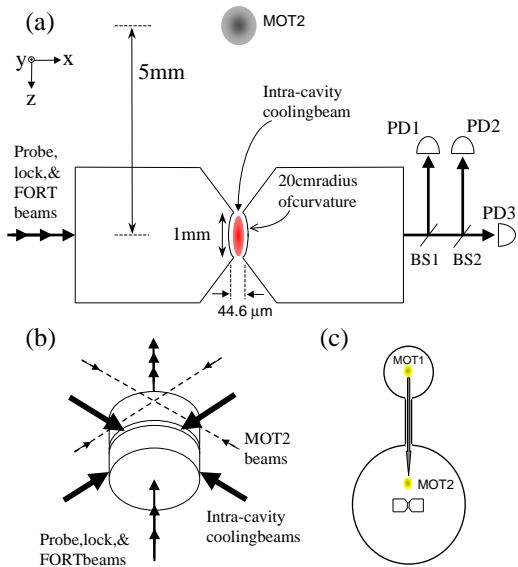}
\caption{Schematic of the experimental apparatus. (a) The dichroic
beam splitter BS1 sends the cavity-length-stabilizing beam to PD1.
BS2 separates the FORT (sent to PD2 for locking) and cavity QED
beams (sent to PD3 for balanced heterodyne detection). (b) Beam
geometry for intra-cavity cooling and MOT-$2$. (c) Differentially
pumped chamber and the two-stage MOT.}
  \end{center}
\end{figure}

\begin{figure}[htbp]
\label{}
  \begin{center}
   \leavevmode
     \epsfxsize=7.4cm  \epsfbox{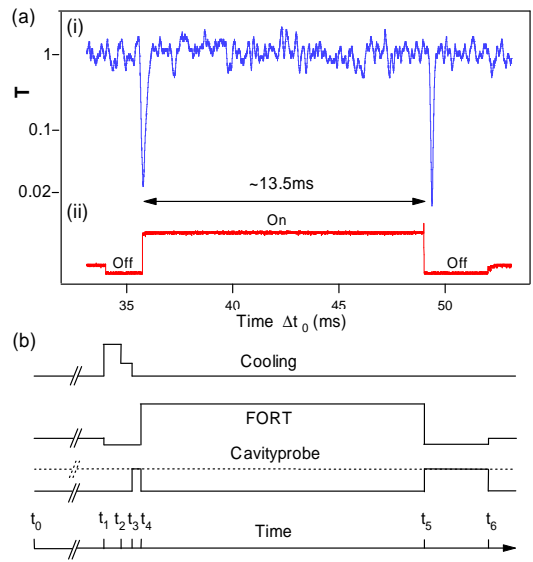}
\caption{(a) A single atom (curve (i), with 30 kHz bandwidth) is
detected within the cavity mode and triggers {\it ON} a
dipole-force trap (FORT, curve (ii)). When the FORT is switched
{\it OFF} after a $13.5$ ms delay, the atom is detected again.
$\Delta _{probe}=0=\Delta _{ac}$, $\Delta _{FORT}^{g}=-45$MHz, and
$\bar{n}=0.1$photons. (b) Timing sequence for the release of
MOT-$2$, intra-cavity cooling, single atom detection, and loading
of the FORT. The polarization-gradient cooled atoms are released
at time $t_{0}$ ($0$ ms). Under the condition of no cooling pulse
and no FORT, atoms reach the cavity mode between $27$ to $37$ ms.
The cooling pulse is switched {\it ON} and the FORT switched {\it
OFF} at $t_{1}$ ($34$ ms). The cooling beams are then detuned
further by $18$MHz and their intensities decreased between $t_{2}$
($35$ ms) and $t_{3}$ ($35.5$ ms). At the end of cooling ($t_{3}$)
the cavity probe is turned {\it ON}. Once a single atom is
detected, for example at $t_{4}$, which varies within
a $3$ ms window, the FORT is switched {\it ON} and the cavity probe {\it OFF}%
. After a predetermined delay, ($t_{5}-t_{4}$), the FORT is turned
{\it OFF} and the cavity probe {\it ON} for detection ($t_{5}$).
At $t_{6}$ everything is reset for the next cycle. For the data
shown in curve (a) (i), the cavity probe was left on continuously
(dashed line in (b)).}
  \end{center}
\end{figure}

\begin{figure}[htbp]
\label{}
  \begin{center}
   \leavevmode
     \epsfxsize=7.4cm  \epsfbox{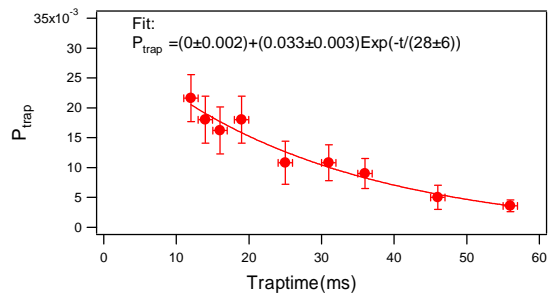}
\caption{Measurement of trap lifetime $\tau _{FORT}$ (see text).
For these data, the FORT is triggered {\it ON} by single atom
transits in the time window $(t_{3},t_{4})$ of Fig. 2(b).
$P_{trap}$ gives the probability per
trigger of successful detection of a trapped atom in the time window $%
(t_{5},t_{6})$. The exponential fit results in $\tau _{FORT}=28\pm
\ 6$ms.}
  \end{center}
\end{figure}

\begin{figure}[htbp]
\label{}
  \begin{center}
   \leavevmode
     \epsfxsize=7.4cm  \epsfbox{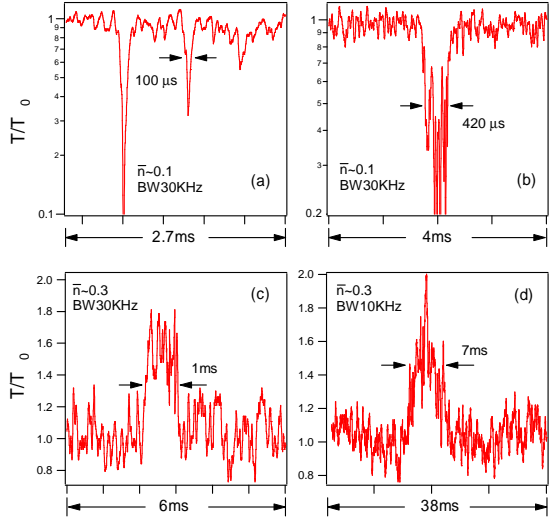}
\caption{Representative transits under four different conditions.
(a) Atom
free falls, $\Delta _{probe}=0=\Delta _{ac}$. (b) Intra-cavity cooling {\it %
ON}, $\Delta _{probe}=0=\Delta _{ac}$ (c) Intra-cavity cooling {\it ON}, $%
\Delta _{probe}=-30$MHz, $\Delta _{ac}=0$. (d) Both intra-cavity
cooling and FORT {\it ON}, $\Delta _{probe}=-10$MHz, $\Delta
_{ac}=-10$MHz, and $\Delta _{FORT}^{g}=-15$MHz.}
  \end{center}
\end{figure}


\begin{references}
\bibitem{birefringence}  The differential phase shift between the cavity
eigen-polarizations is $\Delta \beta =2\mu $rad/bounce.
\end{references}
\end{document}